\documentclass[aps,pre,twocolumn,superscriptaddress,longbibliography]{revtex4-1} 

\usepackage{graphicx}

\begin{document}


\title{Exploring the phase space of multiple states in highly turbulent Taylor-Couette flow}



\author{Roeland C. A. van der Veen}
\author{Sander G. Huisman}
\author{On-Yu Dung}
\author{Ho L. Tang}
\affiliation{Physics of Fluids Group, MESA$^{+}$ Institute  and J.M. Burgers Center for Fluid Dynamics, University of Twente, P.O. Box 217, 7500AE Enschede, The Netherlands}

\author{Chao Sun}
\email{chaosun@tsinghua.edu.cn}
\affiliation{Center for Combustion Energy and Department of Thermal Engineering, Tsinghua University, Beijing 100084, China}
\affiliation{Physics of Fluids Group, MESA$^{+}$ Institute  and J.M. Burgers Center for Fluid Dynamics, University of Twente, P.O. Box 217, 7500AE Enschede, The Netherlands}

\author{Detlef Lohse}
\email{d.lohse@utwente.nl}
\affiliation{Physics of Fluids Group, MESA$^{+}$ Institute  and J.M. Burgers Center for Fluid Dynamics, University of Twente, P.O. Box 217, 7500AE Enschede, The Netherlands}
\affiliation{Max Planck Institute for Dynamics and Self-Organization, 37077 G\"{o}ttingen, Germany}


\date{\today}

\begin{abstract}

We investigate the existence of multiple turbulent states in highly turbulent Taylor-Couette flow in the range of $\mathrm{Ta}=10^{11}$ to $9\cdot10^{12}$, by measuring the global torques and the local velocities while probing the phase space spanned by the rotation rates of the inner and outer cylinder. The multiple states are found to be very robust and are expected to persist beyond $\mathrm{Ta}=10^{13}$. The rotation ratio is the parameter that most strongly controls the transitions between the flow states; the transitional values only weakly depend on the Taylor number. However, complex paths in the phase space are necessary to unlock the full region of multiple states. Lastly, by mapping the flow structures for various rotation ratios in a Taylor-Couette setup with an equal radius ratio but a larger aspect ratio than before, multiple states were again observed. Here, they are characterized by even richer roll structure phenomena, including, for the first time observed in highly turbulent TC flow, an antisymmetrical roll state.

\end{abstract}

\pacs{}

\maketitle

\section{Introduction}

Taylor-Couette (TC) flow, the flow between two coaxial cylinders that can independently rotate, is one of the paradigmatic systems of physics of fluids. In the past century, a large range of phenomena and concepts have been studied in this system, such as instabilities, nonlinear dynamics and spatiotemporal chaos, pattern formation and turbulence. It has been used extensively as a model system in fluid dynamics because it is a closed system, has a relatively simple geometry and therefore has multiple symmetries. This also means that it is experimentally accessible with high precision. For a historical overview of Taylor-Couette research, the reader is referred to \citet{don91}, while the reviews of \citet{pri81} and \citet{far14} treat the rich flow structures of TC flow at the onset of instabilities and slightly above. The state-of-the-art of high Reynolds number Taylor-Couette turbulence is treated by \citefullauthor{gro16} \cite{gro16}, which is the regime we focus on in the present work. Since the research of \citet{wen33} in 1933 and the Austin-Maryland experiments \cite{lat92,lat92a,lew99} in the nineties, only in the past decade more researchers have started to explore the strongly turbulent regime with independently rotating cylinders, both experimentally \cite{ji06,bor10,rav10,pao11,hou11,gil12,hui12,hui14,schartman12,mer13} and numerically \cite{bil07,heTC07,don07,pir08,bra13,ost13,cho14}.

The geometrical parameters of the TC system are the inner and outer cylinder radii $r_i$ and $r_o$ respectively, the gap width $d=r_o-r_i$, and the height of the setup $L$. These can be expressed in dimensionless form by the radius ratio $\eta=r_i/r_o$ and the aspect ratio $\Gamma=L/d$. The inner and outer cylinder rotate with angular velocities $\omega_{i,o}=2\pi f_{i,o}$, which can be expressed in dimensionless form by the Reynolds numbers $\mathrm{Re}_{i,o}=\omega_{i,o}r_{i,o}d/\nu$ where $\nu$ is the kinematic viscosity. Using the analogy of TC flow with Rayleigh-B\'enard (RB) convection~\cite{gro00}, the driving of the flow can alternatively be characterized by the \mbox{Taylor} number \citep{eck07b}
\begin{equation}
\text{Ta}=\frac{(1+\eta)^4}{64\eta^2} \frac{(r_o - r_i)^2 (r_i + r_o)^2 (\omega_i - \omega_o)^2}{\nu^2},
\end{equation}
combined with the (negative) rotation ratio
\begin{equation}
a=-\frac{\omega_o}{\omega_i}
\end{equation}
with $a>0$ for counter-rotation and $a<0$ for corotation. 
The response of the system is the torque $\tau$ required to sustain constant angular velocity or a `Nusselt' number~\cite{eck07b}, $\mathrm{Nu}_\omega = \tau/\tau_\mathrm{laminar}$, which is the angular velocity flux nondimensionalised with the flux of the laminar, nonvortical, flow.

As we increase the driving strength, we can identify three regimes in Taylor-Couette flow. At Taylor numbers that are still low ($\mathrm{Ta}\sim10^6$), the gap between the cylinders is filled with coherent structures called Taylor rolls or Taylor vortices \cite{fen79,and86,far14}. When increasing the driving strength (i.e.~the Taylor number), turbulence starts to develop in the bulk and boundary layers start to develop. These boundary layers are still of laminar type, and in analogy \citep{gro00} with Rayleigh-B\'enard flow, we call this regime the classical regime of TC turbulence. In the turbulent bulk, depending on the rotation ratio $a$, either the Taylor rolls survive, they partly survive close to the inner cylinder, or the bulk is featureless \cite{tok11,cho14,mar14,ost14pd}. By further increasing Ta, around $\mathrm{Ta}\approx3\cdot10^8$, the ultimate regime \citep{kra62,gro11,he12} is reached, in which also the boundary layers are turbulent. It has to be noted that in the transition to the ultimate regime the bulk characteristics do not change significantly, so in principle the Taylor rolls can survive.

The question, however, is what happens to the roll structures for even stronger driving, e.g.\ $\mathrm{Ta}\gg10^{10}$ or $\mathrm{Re}\gg10^5$. Visualizations by \citet{lat92a} show that for pure inner cylinder rotation ($a=0$), no clear Taylor vortices are visible for Reynolds numbers beyond $\mathrm{Re} = 1.2\cdot10^5$. Also for inner cylinder rotation only, \citet{lew99} found that vortex-like structures persist for Reynolds numbers up to $\mathrm{Re} = 10^6$. However, for $\mathrm{Re}>10^5$ the vortex boundaries drifted axially and the number of vortices was not well-defined. Recently, \citet{hui14} explored the existence of rolls at high Reynolds number for counter-rotating cylinders, and found that stable turbulent roll structures persist at a range of rotation ratios $a>0$ (but not $a=0$) for very large Taylor numbers of $\mathrm{Ta}=10^{12}$ ($\mathrm{Re}=10^6$). At pure inner cylinder rotation, these authors found no roll structures in the time-averaged azimuthal velocity, consistent with the previous findings. Strikingly, the range of rotation ratios at which the structures exist, corresponds to the $a$-range close to optimal momentum transport, reflecting that the optimal transport is connected to the existence of the stable large-scale coherent structures: the Taylor rolls supply the optimal angular velocity transfer from the inner to the outer cylinder.

The main finding of \citet{hui14} was however, that at these very high Taylor numbers, multiple states corresponding to different roll structures can exist for the same rotation parameters. These observed multiple states are stable, i.e.~there is no spontaneous switching and there are no slow transitions between the states. The existence of these multiple turbulent states was unexpected and questioned Kolmogorov's paradigm that suggests that for strongly turbulent flows with their many degrees of freedom and large fluctuations, only one turbulent state would be possible \cite{kol41a,kol41b}.

Multiple states have been observed in other geometries, but the nature of the states is generally different. In Rayleigh-B\'enard convection at low Rayleigh number, continuous switching between two different roll states with different heat transfer properties was found \cite{xi08, poe11,wei13}. At larger Rayleigh numbers no multiple states were observed in a single setup; only when the boundary conditions were changed one could trigger transitions \cite{RBchimney,gro11}. However, in \emph{rotating} Rayleigh-B\'enard convection for very high Rayleigh numbers up to $2\cdot10^{12}$, a sequence of sharp transitions was found as the rotation rate was increased \cite{wei15}. For von K\'{a}rm\'{a}n flow multiple turbulent states were found when driving it with impellors with curved blades \cite{Ravelet-PRL2004,Ravelet-JFM2008,Cortet-PRL2010}. These studies revealed the spontaneous symmetry-breaking and turbulent bifurcations in highly turbulent von K\'{a}rm\'{a}n flow up to $\mathrm{Re}=10^6$. Lastly, in spherical-Couette flow spontaneous switching between two turbulent states at fixed rotation rates was observed \cite{zimmerman2011}.

In Taylor-Couette flow, multiple states were only observed in the classical regime (see e.g.\ \cite{and86}) before \citet{hui14} showed their existence in the ultimate regime. More recently, multiple states in TC flow have again been observed, in the transition to the ultimate regime \cite{mar14} and beyond \cite{gul15}. These studies again confirmed that, in the Taylor-Couette geometry, the multiple states manifest themselves as states with a different number of turbulent rolls, which can create a different torque on the cylinders. This leads us to the main implication of the existence of multiple states, which is both fundamental and practical: when extrapolating the known scaling laws for e.g.\ the angular momentum transport, it is essential to be aware of possible transitions between different turbulent states, which in general possess different values for the torque and the angular momentum transport.

In their review of high Reynolds number TC flow, \citefullauthor{gro16} \cite{gro16} posed questions to be solved by the scientific community. Two of them were: ``When further increasing Ta to $\mathrm{Ta}\gg10^{13}$, will the large-scale coherent structures (turbulent Taylor rolls) continue to exist? If so, will multiple turbulent states still coexist, or will the fluctuation be so large that the turbulent dynamics meanders between these states?" These are two questions that we here can begin to answer, by increasing the maximum Taylor number by almost one order of magnitude to $9\cdot10^{12}$ and mapping the (Ta, $a$) phase space of multiple states.

Additionally, as mentioned by \citet{hui14}, understanding of the values of $a$ at which the system transitions between states is still lacking. By exploring the phase spanned by the inner and outer cylinder velocities we will investigate how these transitional $a$ values depend on the driving strength, and what parameter most strongly controls the transitions.

The final question we aim to answer in this work concerns the influence of the aspect ratio of the system. The multiple states manifest themselves as different roll structures and the aspect ratio of the Taylor-Couette system directly influences the number of rolls it can support. In this work we will measure the exact roll structure dependence on the rotation ratio $a$ in a setup with aspect ratio $\Gamma=18.3$, in order to compare the results to previous measurements and to explore the influence of the aspect ratio on the manifestation of multiple states.

We can summarize these questions into one: ``How robust are multiple states?".

  \begin{figure}
 \includegraphics[width=86mm]{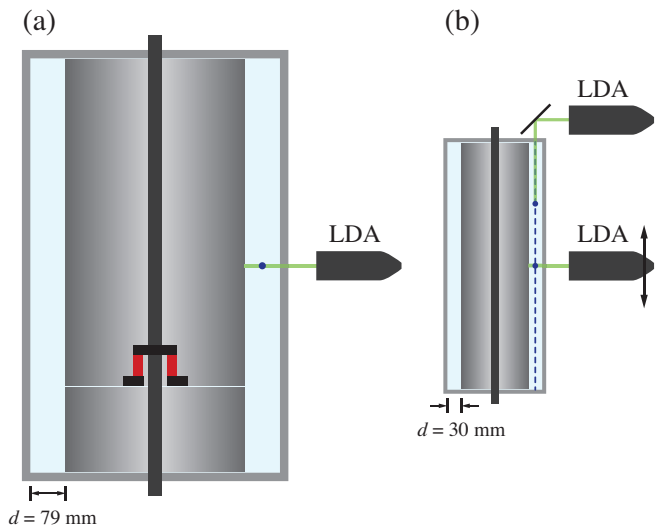}
 \caption{Schematic of the cross section of (a) the T${}^3$C~\cite{gil11a} and (b) the BTTC~\cite{hui15} setup. (a) The T${}^3$C apparatus is outfitted with a co-axial torque transducer to measure the torque that the fluid exerts on the inner cylinder. The azimuthal velocity is probed at the middle of the gap and the middle of the height of the apparatus using laser Doppler anemometry (LDA). (b) In the BTTC apparatus the azimuthal and radial velocity at the middle of the gap and 0.75 of the height are measured through the transparent top plate using a 45$^\circ$ mirror. Additionally, a full height scan of the azimuthal velocity is made by traversing the LDA head vertically.\label{setup}}
 \end{figure}
 
\section{Setups and methods}

The experiments were performed in the Twente Turbulent TC facility (T${}^3$C)~\cite{gil11a} and the new Boiling Twente Taylor-Couette facility (BTTC)~\cite{hui15}; see Fig.~\ref{setup} for a schematic overview. In the larger T${}^3$C setup we are able to reach Taylor numbers of up to $\mathrm{Ta} = 9\cdot10^{12}$ and accurately measure the torque while traversing the phase space of the inner and outer cylinder velocities, in order to map the phase space of multiple turbulent states. The BTTC is designed to study multiphase turbulent flows, but here we will employ the excellent optical access to perform extensive local velocity measurements using laser Doppler anemometry. The systems have equal radius ratio $\eta=r_i/r_o$ but different aspect ratio $\Gamma=L/d$, allowing us to characterize the importance of the aspect ratio with respect to the behavior of the roll structures, by comparing measurements to the previous work of \citet{hui14}. We will first discuss the experimental details of the torque and LDA measurements that are performed in the T${}^3$C facility, after which we will treat the LDA measurements in the BTTC setup. Presently, accurate torque measurements are not possible in the BTTC setup.

The T${}^3$C apparatus has an inner cylinder with radius $r_i = 200\text{ mm}$ and a transparent outer cylinder with inside radius $r_o = 279.4\text{ mm}$ and a height of $L = 927\text{ mm}$. This gives a radius ratio of $\eta = r_i/r_o  = 0.716$ and an aspect ratio of $\Gamma = L/(r_o-r_i) = 11.7$. The maximum rotation rates of the inner and outer cylinder are 20 Hz and 10 Hz, respectively, giving maximum inner and outer Reynolds numbers of $\mathrm{Re}_i = 2.0\cdot10^{6}$ and $\mathrm{Re}_o = 1.4\cdot10^{6}$ with water at room temperature as working fluid. The top and bottom caps are fixed to the outer cylinder. The apparatus was filled with water and actively cooled at the top and bottom plate to keep the temperature constant. Due to excellent turbulent mixing of the fluid, the spatial temperature variation within the system is less than 0.1~K.

\citet{hui14} found that changes between different roll states manifest themselves as `jumps' in the torque that acts on the cylinders. The torque is measured on the middle section of the inner cylinder using a co-axial torque transducer (Honeywell 2404-2K, maximum capacity of 225 Nm). The middle section of the inner cylinder with height $z/L=0.578$ does not cover all the rolls, therefore the exact mean torque value over the entire inner cylinder could be different. This, however, does not take away the jumps observed in the torque when the flow undergoes transitions between different roll structures.

The azimuthal velocity is obtained by laser Doppler anemometry (LDA), see Fig.~\ref{setup}(a). The laser beams go through the outer cylinder and are focused in the middle of the gap.  Curvature effects of the outer cylinder on our LDA system are accounted for by numerically ray-tracing the LDA-beams \cite{hui12b}. The water is seeded with 5 $\mu\mathrm{m}$ diameter polyamide tracer particles (Dantec) with a maximum Stokes number of $\mathrm{St} = \tau_p/\tau_K = 0.017\ll1$ (with $\tau_p$ the particle response time and $\tau_K$ the Kolmogorov timescale), so that they can be considered to be perfect tracers of the flow.

  \begin{figure}
 \includegraphics[scale=0.5]{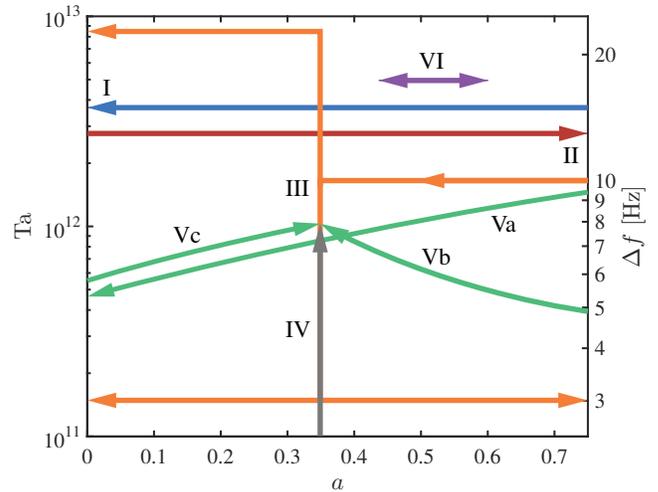}
 \caption{Representative examples of the types of measurement trajectories in the phase space (Ta, $a$): constant Ta with decreasing $a=-f_o/f_i$ (I) and increasing $a$ (II), constant Ta with a change to different Ta in between (III), increasing Ta at constant $a$ (IV), fixed $f_i$ with decreasing $|f_o|$ (Va), fixed $f_o$ with increasing $|f_i|$ (Vb), fixed $f_i$ with increasing $|f_o|$ (Vc) and constant Ta back and forth (VI). For the right axis a temperature of $22^\circ$C is used to convert from the Taylor number to $\Delta f=f_i-f_o$. \label{phasespace0}}
 \end{figure}
 
To answer the questions we posed in the introduction, we measured the torque on the cylinders while quasistatically traversing the phase space along different types of trajectories shown in Fig.~\ref{phasespace0}. Specific trajectories such as those of type III were used to find the extent of the phase space in which multiple states are possible, as will be shown in the results section. 

We slowly followed the trajectories in phase space by changing the cylinder velocities with 4Hz/h, found to be slow enough to make acceleration/deceleration effects negligible and give the system ample time to stabilize into any flow state \cite{gil11,hui14}. The beginnings of the trajectories were reached by spinning the cylinders up to the initial velocities at constant $a$ with 4Hz/min, after which the system was kept at these velocities for five minutes to remove any transitional effects. The mean water temperature for each run was between $20^\circ$C to $27^\circ$C, while the temperature variation during each measurement was smaller than 0.5 K. The Taylor number depends on viscosity and thus on temperature; we therefore removed the main temperature dependence by compensating the Nusselt number with Ta$^{0.4}$, because $\mathrm{Nu}_{\omega} \propto \mathrm{Ta}^{0.4}$ in the present parameter regime. This approach has been followed before, see for example Refs.~\cite{gil11,pao11,hui14}.

We now turn our attention to the BTTC facility~\cite{hui15}. This setup has an inner cylinder with radius $r_i = 75\text{ mm}$ and a transparent outer cylinder with inside radius $r_o = 105\text{ mm}$, and a height of $L = 549\text{ mm}$. This gives a radius ratio of $\eta = r_i/r_o  = 0.714$, very close to that of the T${}^3$C setup, and an aspect ratio of $\Gamma = L/(r_o-r_i) = 18.3$, considerably larger than that of the T${}^3$C setup. The maximum rotation rates of the inner and outer cylinder are again 20 Hz and 10 Hz, respectively, giving maximum inner and outer Reynolds numbers of $\mathrm{Re}_i = 2.8\cdot10^{5}$ and $\mathrm{Re}_o = 2.0\cdot10^{5}$ with water at room temperature as the working fluid. The top and bottom caps are fixed to the outer cylinder. The top plate and outer cylinder are fully transparent, enabling excellent optical access for optical measurement techniques such as laser Doppler anemometry. The apparatus was filled with water and actively cooled at the inner cylinder to keep the temperature constant. The spatial temperature variation within the system is less than 0.1~K.


Two types of LDA measurements were performed in the BTTC setup. The first consisted of measuring the local velocity at a single point continuously while slowly changing $a$, in order to see transitions between the multiple states. With the second set of experiments we characterized the roll structures by performing finely spaced measurements of the velocity over the height of the setup, for several values of $a$. For each type of measurement, a different way of mounting the LDA system was employed, as shown in Fig.~\ref{setup}(b). Using a mirror mounted at an angle of 45$^\circ$, we could measure the azimuthal and radial velocity in the flow. Although measuring the azimuthal velocity is a good way of characterizing rolls \cite{hui14}, measuring the radial velocity gives a more direct quantification of the rolls. We quasistatically traversed the phase space by either decreasing or increasing $a$ while keeping the difference in cylinder velocities constant (and so Ta), while simultaneously measuring the two velocity components with LDA at the middle of the gap and at $z/L=0.75$, showing transitions between different roll states. This was done at $\Delta f = f_i-f_o=10$~Hz, giving a Taylor number of $\mathrm{Ta}=3\cdot10^{10}$.  Again 5 $\mu\mathrm{m}$ diameter polyamide tracer particles were used, which here have a maximum Stokes number of $\mathrm{St} = \tau_p/\tau_\eta = 0.002\ll1$.

In addition to these continuous measurements of the velocity, we performed axial scans using LDA. The LDA head was positioned on a vertical translation stage with the focus at the middle of the gap (optical effects were again accounted for using a ray-tracer). The measurements were performed in a similar manner as by \citet{hui14}, and can be summarized as follows: we spun the cylinders up to either $a=0$ or $a=0.75$, acquired velocity data at many points along the height of the cylinder, slowly changed the velocity to the next value of $a$, repeated the velocity acquisition, again slowly changed to the next $a$, and so forth. At every height the velocity was measured for 20 s, giving approximately 10000 data points each. The axial step size was 1.5 mm and the step size in $a$ was 0.05, refined to 0.025 around $a=0.4$. The measurement height was not traversed monotonically, but divided in groups of heights, which were measured non-consecutively. We did so in order to detect possible slow transitions of the flow, which then would show as discontinuities in the resulting velocity profiles. 

 \begin{figure}
 \includegraphics[scale=0.5]{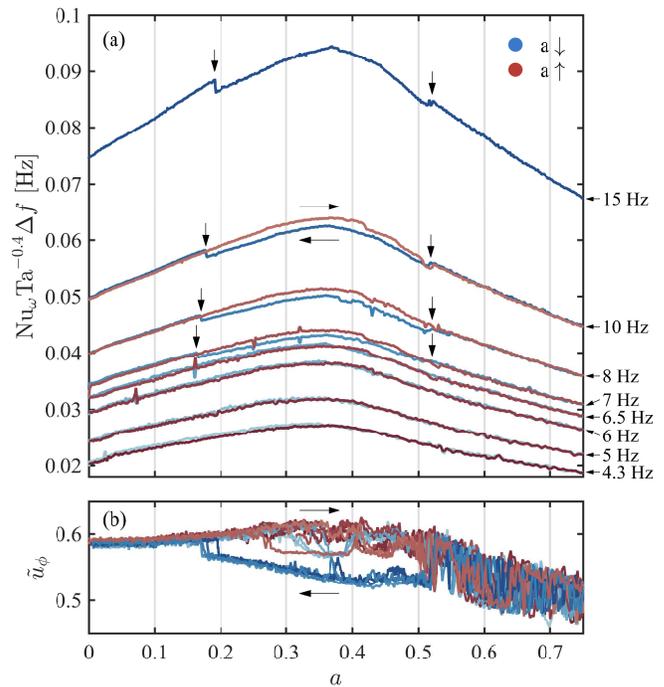}
 \caption{Global torque and local velocity versus rotation ratio for several values of Ta, corresponding to trajectories of type I (blue) and II (red) in Fig.~\ref{phasespace0}. (a) Compensated $\textrm{Nu}_{\omega}$ as a function of $a$, multiplied by $\Delta f = f_i-f_o$ to shift the lines for clarity, for $\Delta f = 4.3$ Hz, 5 Hz, 6 Hz, 6.5 Hz, 7 Hz, 8 Hz, 10 Hz and 15 Hz (as indicated on the right). The vertical arrows denote the transitions to the low and high torque states. (b) Azimuthal velocity $\tilde{u}_\phi=(\langle u_\phi \rangle_t - u_i)/(u_o-u_i)$ measured at $r=(r_i+r_o)/2$ and $z/L=0.5$ as a function of $a$. \label{jumps}}
 \end{figure}

\section{Phase space of multiple states in the T$^3$C setup with $\Gamma=11.7$}

As observed before \cite{hui14}, different flow states can be triggered by either decreasing (trajectory type I) or increasing (trajectory type II) the rotation ratio $a$ at constant driving parameter Ta. In this paper we performed these measurements at several different values of the Taylor number between $\mathrm{Ta}=1.4\cdot10^{11}$ and  $\mathrm{Ta}=8.8\cdot10^{12}$, of which some examples are shown in Fig.~\ref{jumps}(a). For the case of increasing $a$, the torque is continuous and shows the familiar peak around $a=0.36$ \cite{gil11,pao11,gil12,mer13,ost14pd}. But when $a$ is decreased, the system can enter a state with a lower torque (called `low state') at a certain value $a>0.36$, jumping back to a higher torque state (`high state') at a certain value $a<0.36$. This bifurcation is confirmed by measuring the local velocity using laser Doppler anemometry, see Fig.~\ref{jumps}(b). The transition to the low state is a probabilistic phenomenon, although the probability is high (8 out of 10 for trajectories at $\Delta f = 8$ Hz).

By performing several more measurements with finely spaced Taylor number it was found that for the type I and II trajectories of constant Ta, multiple states only appear for $\mathrm{Ta}>(7.7\pm0.2) \cdot10^{11}$. We stress that this does not imply anything about the existence or strength of roll structures, but rather the ability of the system to switch between the different roll structures. Of the two possible roll states, the 8-roll state corresponding to the high torque state seems to be more favorable than the 6-roll state corresponding to the low torque state \cite{hui14}.

From these measurements we can draw two intermediate conclusions, on which we will then further build. The first is that multiple states still occur for higher Ta than observed before, but also only beyond a certain high enough threshold Taylor number. The second is that there seems to be a (weak) dependence of the transitional $a$-values on the Taylor number.

  \begin{figure}
 \includegraphics[scale=0.5]{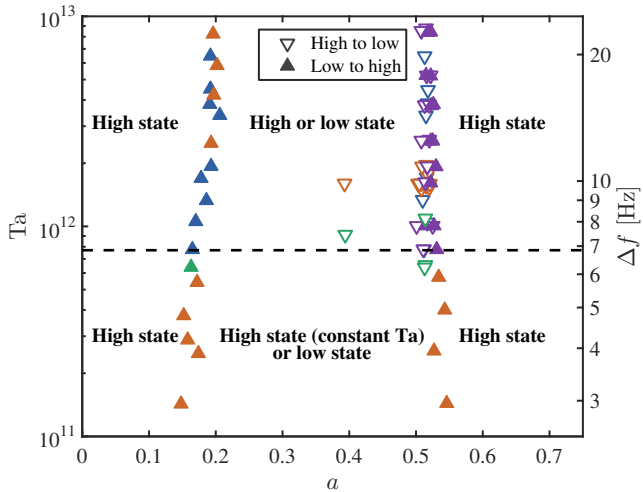}
 \caption{The phase space (Ta,a) showing transitions from the high to low state by open symbols and transitions from the low to high state by closed symbols. The dashed line is at $\mathrm{Ta}=7.7 \cdot10^{11}$, below which the system does not transition into the low state for trajectories of type I, II and VI. The colors correspond to those in Fig.~\ref{phasespace0}. \label{phasespace}}
 \end{figure}

Refer now to Fig.~\ref{phasespace}, as we make a journey through the trajectories and fill the phase space with the transitions between the low and high torque state. We have already seen that at constant Ta and increasing $a$ no transitions are present, but for decreasing $a$ we can plot the blue symbols in the figure. As mentioned, below $\mathrm{Ta}=(7.7\pm0.2) \cdot10^{11}$ the low state cannot be reached using these trajectories. We find that by following a type IV trajectory of constant $a$ and increasing Ta, the system sometimes is in the low state, however with a low probability \cite{hui14}. To further explore the boundaries of the multiple states, we follow the recipe of the type III trajectories, by starting at $a=1$ and $\Delta f = 10$ Hz, then reducing $a$ at constant Ta to relatively reliably enter the low state. Then Ta is reduced at constant $a$, crossing the line of $\mathrm{Ta}=7.7 \cdot10^{11}$, after which $a$ is either decreased or increased. The system stays in the low state until jumping back to the high state at $a=0.15$ to 0.17 and $a=0.53$ to 0.55, respectively. This greatly extends the knowledge of the boundaries of the region where multiple states are possible. To gather more points at the left boundary of the phase space, the Taylor number is also increased to $\mathrm{Ta}=8.8\cdot10^{12}$, corresponding to a combined Reynolds number \cite{gil11a} of $\mathrm{Re}=(\omega_i r_i-\omega_o r_o)d/\nu = 2.6\cdot10^6$. Concluding, multiple states are observed for almost two orders of magnitude in the Taylor number within a well-defined range of rotation ratios.

  \begin{figure}
 \includegraphics[scale=0.5]{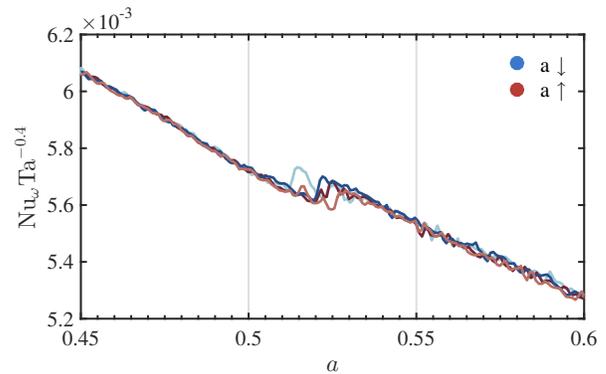}
 \caption{Global torque signal of one measurement following a type IV trajectory at $\Delta f =15$ Hz, decreasing and increasing $a$ twice, showing hysteresis in the transitional $a$-value. The duration of this part of the measurement trajectory, from $a=0.45$ to $0.60$ (and vice versa), is 14 minutes and 33 seconds.\label{hysteresis}}
 \end{figure}

We wonder whether the transitional $a$ depends on the way one traverses the boundary of the multiple states. For this, we perform experiments at constant $f_i$ or $f_o$, traversing the (Ta, $a$) phase space diagonally. Along trajectories of type Va the system undergoes transitions at very similar $a$ as compared to the other trajectories, confirming the idea that the rotation ratio is the parameter that controls the inclination of the system to be in either state. By increasing $|f_i|$ at constant $f_o$ (trajectory type Vb) the system also undergoes transitions to the low state at the expected value of $a\approx0.51$ (except for 1 out of 5 measurements). This transition however, is incidentally at a lower $\mathrm{Ta}=6.4 \cdot10^{11}$ than the previously observed boundary of $\mathrm{Ta}=(7.7\pm0.2) \cdot10^{11}$ for trajectories of constant Ta.

So far we have approached the right boundary in the phase space only from the high state. By now first traversing it from the right and then reversing the direction of $a$, we can investigate the system jumping back to the high state: does hysteresis of the transition position occur and can we quantify it? This was done by following trajectories of type VI at several different Taylor numbers. In the phase space we can see that for most of the measurements the systems undergoes transitions back to the high state slightly later (i.e.~at higher $a$) when increasing $a$ again. In Fig.~\ref{hysteresis} an example of a measurement is shown where the direction of $a$ is changed three times. The difference in the value of $a$ between the jumps back and forth was less than 0.015 here, while in general it was smaller than 0.030. We conclude that there is some hysteresis, which is only just revealed within the variation of the jump positions between measurements.

Having filled the phase space, we can now see how the boundaries of the multiple states region behave. The previously found range of $0.17<a<0.51$ holds for a large range of Ta, and only widens slightly with decreasing Ta below $\mathrm{Ta}=10^{12}$. When looking at the larger values of the Taylor number, there is no indication of the multiple states region closing up. From this, we expect the multiple states to exist (far) beyond $\mathrm{Ta}=10^{13}$.

\section{Flow structures in the BTTC setup with $\Gamma=18.3$}

  \begin{figure}
 \includegraphics[scale=1]{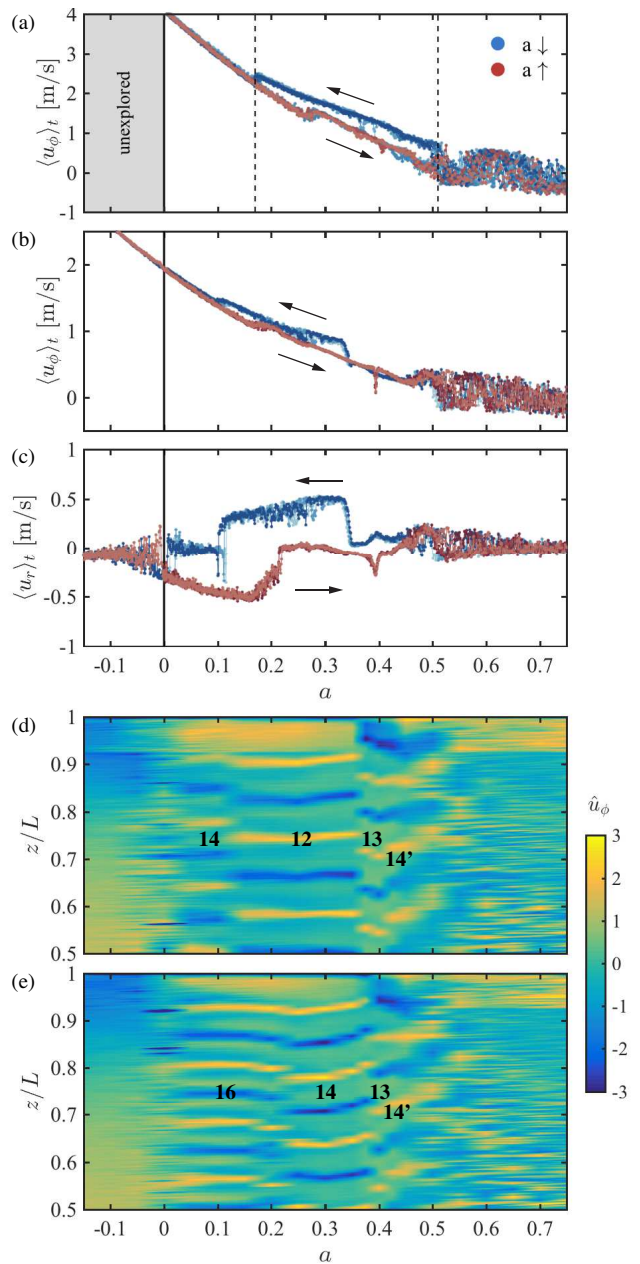}
 \caption{Local velocity and flow structures. (a) Azimuthal velocity measured in the T${}^3$C setup at $r=(r_i+r_o)/2$ and $z/L=0.5$ as a function of $a$ for increasing and decreasing $a$ at $\mathrm{Ta}\approx10^{12}$ (data from \cite{hui14}). (b) and (c) Azimuthal and radial velocity, respectively, in the BTTC setup at $r=(r_i+r_o)/2$ and $z/L=0.75$ as a function of $a$ for increasing and decreasing $a$ at $\mathrm{Ta}=3\cdot10^{10}$. (d) and (e) Axial scans of the standardized azimuthal velocity in the BTTC setup as a function of $a$ for decreasing (d) and increasing (e) $a$ at $\mathrm{Ta}=3\cdot10^{10}$. The standardized velocity is $\hat{u}_\phi=(\langle u_\phi \rangle_t-\langle u_\phi \rangle_z)/\sigma_a(u_\phi)$ where $\sigma_a(u_\phi)$ is the standard deviation of $u_\phi$ for each a, and $\langle \rangle_x$ stands for averaging over $x$. The bold numbers indicate the number of rolls the systems contains over the full height, with the accent denoting a reversed roll direction. \label{states}}
 \end{figure}

Having shown the dependence of the existence of multiple states on the rotation ratio and driving strength, we will now turn our attention to another very important parameter: the aspect ratio $\Gamma$ of the system. It is straightforward to see that the aspect ratio will strongly influence the roll structures of the system. But what do the roll structures corresponding to the different states actually look like? That is what we aim to answer in this section. \citet{hui14} have mapped the roll structures for increasing and decreasing $a$ in the T${}^3$C setup, showing that 8 and 6-roll states can exist for an aspect ratio of 11.7, with the 6-roll state corresponding to the non-default `low' torque state. Here we performed similar experiments in the BTTC setup with an aspect ratio of 18.3, with a finer axial resolution to individually resolve the expected larger number of rolls.

First, let us look at measurements of the azimuthal velocity at a single positions for both setups, see Fig.~\ref{states}(a) and (b). In both cases the velocity bifurcates, showing very stable multiple states. For $a>0.51$ the velocity fluctuates strongly, which is not suggestive of a stable, well-defined flow state. However, the velocity was measured at the middle of the gap, and it could be possible that coherent structures still existed closer to the inner cylinder \cite{ost13}. Now, as can be seen in Fig.~\ref{states}(c), the radial velocity is a stronger indicator of the multiple states than the azimuthal velocity. While only two branches were observed, one for increasing $a$ and one for decreasing $a$, there are strong jumps of the velocity within these branches. This indicates that more than one state is possible in either branch.

By performing the aforementioned axial scans of the azimuthal velocity, we saw that this indeed is the case. Fig.~\ref{states}(d) shows that for $a$ outside [0,0.5] there are no clearly defined structures, but that within that range, a multitude of states is accessible. To count the number of rolls, we used the following concept: in between a pair of counter-rotating rolls, either high-velocity fluid from the inner cylinder or low-velocity fluid from the outer cylinder is advected, creating maxima and minima in the standardized velocity. From this we deduced that one roll sits between every minimum and maximum. With this knowledge, we counted 12, 13, and 14 rolls for different regions of $a$, for the measurement with decreasing $a$.

The case of increasing $a$, Fig.~\ref{states}(e), also shows a myriad of accessible states, with 13, 14 and 16 rolls. Curiously, for $a$ within the range of [0,0.4] the two branches of increasing and decreasing $a$ are never in the same roll state, again showing the importance of the history of the system. By following different trajectories in phase space, one can end up in different states, which are stable for at least multiple hours. Apparently, information about the history of the flow is not destroyed by the strong turbulence.

The average aspect ratio of the observed rolls were 1.53, 1.41, 1.31 and 1.14 for 12, 13, 14 and 16 rolls, respectively. The aspect ratios of the rolls observed in the T${}^3$C with smaller system aspect ratio of $\Gamma=11.7$ were 1.96 and 1.46, measured however at a larger Taylor number of $\mathrm{Ta}=10^{12}$. The difference in aspect ratio can be explained by a trend of increasing aspect ratio with Reynolds number (and Taylor number) \cite{cho14}, although relatively stronger end effects in the T${}^3$C setup with smaller $\Gamma$ could also play a role. The fact that more roll states can be observed in the BTTC setup can be explained by the smaller energy barrier the system has to overcome to change the roll state. When adding or removing a single roll in a system with large aspect ratio $\Gamma$, the average roll aspect ratio changes less than for a system with smaller aspect ratio $\Gamma$.

  \begin{figure}[t!]
 \includegraphics[scale=0.5]{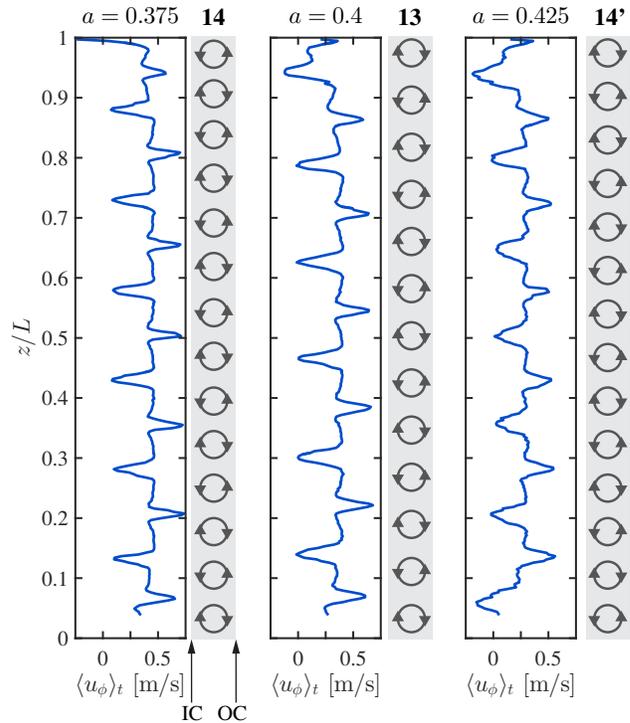}
 \caption{Azimuthal velocity profiles at $\mathrm{Ta}=3\cdot10^{10}$ as a function of height $z/L$ for $a=0.375$, $a=0.4$ and $a=0.425$, with increasing $a$. The data are from Fig.~\ref{states}(e) but here shown for the full height of the system. The gray circles with arrows schematically show the rolls and their flow directions. Here the inner cylinder (IC) is on the left, the outer cylinder (OC) is on the right, and the convention of positive IC rotation is used.\label{profiles}}
 \end{figure}

Signs of symmetry-breaking appear in Fig.~\ref{states}(e) for $a$ in the range [0.2,0.4], where the top roll slowly changes aspect ratio, previously also observed in the T${}^3$C setup. More strikingly, at $a=0.4$ a roll state with an odd number of rolls seems to have appeared, which is very unexpected considering the symmetric boundary conditions of the system. We will now look into this phenomenom more closely using Fig.~\ref{profiles}. By comparing the case of $a=0.375$ and $a=0.425$, we found that the system was in the same roll state of 14 vortices for these values of $a$, however with reversed flow direction of the rolls. We remind the reader that in this setup the end plates rotate with the outer cylinder, and it can be expected that the preferred roll direction at the boundaries depends on the rotation ratio $a$. In between these two cases, at $a=0.4$, both states are seemingly not favorable and the system reacts by first changing (in this particular case) the flow direction of the top roll of the system to the new boundary condition. Now that the top and bottom rolls rotate in the same direction, only an odd number of rolls can be supported in the system. This state has the center of the middle roll at $z/L=0.5$, making the term `antisymmetrical' appropriate. Moving on to $a=0.425$, also the bottom roll in the system has adapted, arriving again at the expected symmetrical roll structure. In the case of decreasing $a$, the odd 13-roll state can also be observed, although the system does not revert to the 14-roll state until much later (smaller) $a$. 

To our knowledge this is the first time a stable antisymmetrical strongly turbulent roll structure has been observed in a symmetrical Taylor-Couette setup. More generally, symmetry-breaking has been observed before in von K\'{a}rm\'{a}n flow with curved propellors \cite{caballero2013}. However, in that setup it is directly caused by symmetry breaking of the boundary conditions due to the curved blades, while in our TC setup the top and bottom end plates are symmetrical. Additionally, in von K\'{a}rm\'{a}n flow the system jumps randomly and spontaneously between the two asymmetrical states, while the symmetry breaking observed here is stable.

\section{Conclusions}
To answer the question we ended the introduction with: multiple states in highly turbulent Taylor-Couette flow are very robust. In this study their existence was observed for a Taylor number range of almost two decades, and it can be expected that they persist for $\mathrm{Ta}=10^{13}$ and beyond. Although their existence is robust, to reliably reach the less-likely low torque state in our T${}^3$C setup, complex paths in phase space can be necessary. When a path of constant Ta (i.e.~constant difference of rotation rates) is taken, a minimum Taylor number of $\mathrm{Ta}=(7.7\pm0.2) \cdot10^{11}$ is needed to observe multiple states.

The rotation ratio $a$ is found to be the main control parameter for the transition; only a very weak dependence of the transitional $a$-values on the Taylor number was observed. Traversing the boundaries of the phase space with non-constant $a$ and Ta also does not significantly  influence the transitional $a$ value.

The robustness also reflects in the measurements we performed in a different setup with a larger aspect ratio. Here, the phenomena are even richer, although the same principles hold, namely that the system is hysteretic and can be in multiple stable turbulent states for the same driving parameters. For the larger aspect ratio, the system can be in a larger number of different roll states. Additionally, for the first time in turbulent TC flow, an antisymmetrical roll state with an odd number of rolls has been observed.

A theoretical understanding of the values of $a$ at which the systems transitions between states remains elusive, although we have shown here that the transitional $a$ values strongly depend on the aspect ratio of the system. Another interesting opportunity for further research is the probabilistic nature of the bifurcations.

With this work we have continued the investigation into the manifestation and selectability of coherent states in highly turbulent flows, which is important for understanding and predicting large-scale flows in nature and industry.

\begin{acknowledgments}
We would like to thank M. Bos, G.-W. Bruggert and G. Mentink for their technical support and R. Ezeta and R. Verschoof for useful discussions. This work was financially supported by a European Research Council (ERC) Advanced Grant, the Simon Stevin Prize of the Technology Foundation STW of The Netherlands, and European High-Performance Infrastructures in Turbulence (EuHIT).
\end{acknowledgments}

\bibliography{MS2_Veen.bib}

\begin{thebibliography}{52}%
\makeatletter
\providecommand \@ifxundefined [1]{%
 \@ifx{#1\undefined}
}%
\providecommand \@ifnum [1]{%
 \ifnum #1\expandafter \@firstoftwo
 \else \expandafter \@secondoftwo
 \fi
}%
\providecommand \@ifx [1]{%
 \ifx #1\expandafter \@firstoftwo
 \else \expandafter \@secondoftwo
 \fi
}%
\providecommand \natexlab [1]{#1}%
\providecommand \enquote  [1]{``#1''}%
\providecommand \bibnamefont  [1]{#1}%
\providecommand \bibfnamefont [1]{#1}%
\providecommand \citenamefont [1]{#1}%
\providecommand \href@noop [0]{\@secondoftwo}%
\providecommand \href [0]{\begingroup \@sanitize@url \@href}%
\providecommand \@href[1]{\@@startlink{#1}\@@href}%
\providecommand \@@href[1]{\endgroup#1\@@endlink}%
\providecommand \@sanitize@url [0]{\catcode `\\12\catcode `\$12\catcode
  `\&12\catcode `\#12\catcode `\^12\catcode `\_12\catcode `\%12\relax}%
\providecommand \@@startlink[1]{}%
\providecommand \@@endlink[0]{}%
\providecommand \url  [0]{\begingroup\@sanitize@url \@url }%
\providecommand \@url [1]{\endgroup\@href {#1}{\urlprefix }}%
\providecommand \urlprefix  [0]{URL }%
\providecommand \Eprint [0]{\href }%
\providecommand \doibase [0]{http://dx.doi.org/}%
\providecommand \selectlanguage [0]{\@gobble}%
\providecommand \bibinfo  [0]{\@secondoftwo}%
\providecommand \bibfield  [0]{\@secondoftwo}%
\providecommand \translation [1]{[#1]}%
\providecommand \BibitemOpen [0]{}%
\providecommand \bibitemStop [0]{}%
\providecommand \bibitemNoStop [0]{.\EOS\space}%
\providecommand \EOS [0]{\spacefactor3000\relax}%
\providecommand \BibitemShut  [1]{\csname bibitem#1\endcsname}%
\let\auto@bib@innerbib\@empty
\bibitem [{\citenamefont {Donnelly}(1991)}]{don91}%
  \BibitemOpen
  \bibfield  {author} {\bibinfo {author} {\bibfnamefont {R.~J.}\ \bibnamefont
  {Donnelly}},\ }\bibfield  {title} {\enquote {\bibinfo {title}
  {{{Taylor-Couette}} flow: the early days},}\ }\href@noop {} {\bibfield
  {journal} {\bibinfo  {journal} {Phys. Today}\ }\textbf {\bibinfo {volume}
  {44(11)}},\ \bibinfo {pages} {32--39} (\bibinfo {year} {1991})}\BibitemShut
  {NoStop}%
\bibitem [{\citenamefont {Di~Prima}\ and\ \citenamefont
  {Swinney}(1985)}]{pri81}%
  \BibitemOpen
  \bibfield  {author} {\bibinfo {author} {\bibfnamefont {R.~C.}\ \bibnamefont
  {Di~Prima}}\ and\ \bibinfo {author} {\bibfnamefont {H.~L.}\ \bibnamefont
  {Swinney}},\ }\href@noop {} {\emph {\bibinfo {title} {Hydrodynamic
  Instabilities and the Transition to Turbulence}}},\ edited by\ \bibinfo
  {editor} {\bibfnamefont {H.~L.}\ \bibnamefont {Swinney}}\ and\ \bibinfo
  {editor} {\bibfnamefont {J.~P.}\ \bibnamefont {Gollub}},\ \bibinfo {series}
  {Topics in Applied Physics}, Vol.~\bibinfo {volume} {45}\ (\bibinfo
  {publisher} {Springer},\ \bibinfo {year} {1985})\ pp.\ \bibinfo {pages}
  {139--180}\BibitemShut {NoStop}%
\bibitem [{\citenamefont {Fardin}\ \emph {et~al.}(2014)\citenamefont {Fardin},
  \citenamefont {Perge},\ and\ \citenamefont {Taberlet}}]{far14}%
  \BibitemOpen
  \bibfield  {author} {\bibinfo {author} {\bibfnamefont {M.~A.}\ \bibnamefont
  {Fardin}}, \bibinfo {author} {\bibfnamefont {C.}~\bibnamefont {Perge}}, \
  and\ \bibinfo {author} {\bibfnamefont {N.}~\bibnamefont {Taberlet}},\
  }\bibfield  {title} {\enquote {\bibinfo {title} {{``}{{The}} hydrogen atom of
  fluid dynamics{"} - {{Introduction}} to the {{Taylor-Couette}} flow for
  {{Soft Matter}} scientists},}\ }\href@noop {} {\bibfield  {journal} {\bibinfo
   {journal} {Soft Matter}\ }\textbf {\bibinfo {volume} {10}},\ \bibinfo
  {pages} {3523--3535} (\bibinfo {year} {2014})}\BibitemShut {NoStop}%
\bibitem [{\citenamefont {Grossmann}\ \emph {et~al.}(2016)\citenamefont
  {Grossmann}, \citenamefont {Lohse},\ and\ \citenamefont {Sun}}]{gro16}%
  \BibitemOpen
  \bibfield  {author} {\bibinfo {author} {\bibfnamefont {S.}~\bibnamefont
  {Grossmann}}, \bibinfo {author} {\bibfnamefont {D.}~\bibnamefont {Lohse}}, \
  and\ \bibinfo {author} {\bibfnamefont {C.}~\bibnamefont {Sun}},\ }\bibfield
  {title} {\enquote {\bibinfo {title} {High {{Reynolds}} number
  {{Taylor-Couette}} turbulence},}\ }\href@noop {} {\bibfield  {journal}
  {\bibinfo  {journal} {Annu. Rev. Fluid Mech.}\ }\textbf {\bibinfo {volume}
  {48}},\ \bibinfo {pages} {53--80} (\bibinfo {year} {2016})}\BibitemShut
  {NoStop}%
\bibitem [{\citenamefont {Wendt}(1933)}]{wen33}%
  \BibitemOpen
  \bibfield  {author} {\bibinfo {author} {\bibfnamefont {F.}~\bibnamefont
  {Wendt}},\ }\bibfield  {title} {\enquote {\bibinfo {title} {Turbulente
  {{Str\"omungen}} zwischen zwei rotierenden konaxialen {{Zylindern}}},}\
  }\href@noop {} {\bibfield  {journal} {\bibinfo  {journal} {Ingenieur-Archiv}\
  }\textbf {\bibinfo {volume} {4}},\ \bibinfo {pages} {577--595} (\bibinfo
  {year} {1933})}\BibitemShut {NoStop}%
\bibitem [{\citenamefont {Lathrop}\ \emph
  {et~al.}(1992{\natexlab{a}})\citenamefont {Lathrop}, \citenamefont
  {Fineberg},\ and\ \citenamefont {Swinney}}]{lat92}%
  \BibitemOpen
  \bibfield  {author} {\bibinfo {author} {\bibfnamefont {D.~P.}\ \bibnamefont
  {Lathrop}}, \bibinfo {author} {\bibfnamefont {J.}~\bibnamefont {Fineberg}}, \
  and\ \bibinfo {author} {\bibfnamefont {H.~L.}\ \bibnamefont {Swinney}},\
  }\bibfield  {title} {\enquote {\bibinfo {title} {Turbulent flow between
  concentric rotating cylinders at large {{Reynolds}} numbers},}\ }\href@noop
  {} {\bibfield  {journal} {\bibinfo  {journal} {Phys. Rev. Lett.}\ }\textbf
  {\bibinfo {volume} {68}},\ \bibinfo {pages} {1515--1518} (\bibinfo {year}
  {1992}{\natexlab{a}})}\BibitemShut {NoStop}%
\bibitem [{\citenamefont {Lathrop}\ \emph
  {et~al.}(1992{\natexlab{b}})\citenamefont {Lathrop}, \citenamefont
  {Fineberg},\ and\ \citenamefont {Swinney}}]{lat92a}%
  \BibitemOpen
  \bibfield  {author} {\bibinfo {author} {\bibfnamefont {D.~P.}\ \bibnamefont
  {Lathrop}}, \bibinfo {author} {\bibfnamefont {J.}~\bibnamefont {Fineberg}}, \
  and\ \bibinfo {author} {\bibfnamefont {H.~L.}\ \bibnamefont {Swinney}},\
  }\bibfield  {title} {\enquote {\bibinfo {title} {Transition to shear-driven
  turbulence in {{Couette-Taylor}} flow},}\ }\href@noop {} {\bibfield
  {journal} {\bibinfo  {journal} {Phys. Rev. A}\ }\textbf {\bibinfo {volume}
  {46}},\ \bibinfo {pages} {6390--6405} (\bibinfo {year}
  {1992}{\natexlab{b}})}\BibitemShut {NoStop}%
\bibitem [{\citenamefont {Lewis}\ and\ \citenamefont {Swinney}(1999)}]{lew99}%
  \BibitemOpen
  \bibfield  {author} {\bibinfo {author} {\bibfnamefont {G.~S.}\ \bibnamefont
  {Lewis}}\ and\ \bibinfo {author} {\bibfnamefont {H.~L.}\ \bibnamefont
  {Swinney}},\ }\bibfield  {title} {\enquote {\bibinfo {title} {Velocity
  structure functions, scaling, and transitions in high-{{Reynolds}}-number
  {{Couette-Taylor}} flow},}\ }\href@noop {} {\bibfield  {journal} {\bibinfo
  {journal} {Phys. Rev. E}\ }\textbf {\bibinfo {volume} {59}},\ \bibinfo
  {pages} {5457--5467} (\bibinfo {year} {1999})}\BibitemShut {NoStop}%
\bibitem [{\citenamefont {Ji}\ \emph {et~al.}(2006)\citenamefont {Ji},
  \citenamefont {Burin}, \citenamefont {Schartman},\ and\ \citenamefont
  {Goodman}}]{ji06}%
  \BibitemOpen
  \bibfield  {author} {\bibinfo {author} {\bibfnamefont {H.}~\bibnamefont
  {Ji}}, \bibinfo {author} {\bibfnamefont {M.}~\bibnamefont {Burin}}, \bibinfo
  {author} {\bibfnamefont {E.}~\bibnamefont {Schartman}}, \ and\ \bibinfo
  {author} {\bibfnamefont {J.}~\bibnamefont {Goodman}},\ }\bibfield  {title}
  {\enquote {\bibinfo {title} {{Hydrodynamic turbulence cannot transport
  angular momentum effectively in astrophysical disks}},}\ }\href@noop {}
  {\bibfield  {journal} {\bibinfo  {journal} {Nature}\ }\textbf {\bibinfo
  {volume} {444}},\ \bibinfo {pages} {343--346} (\bibinfo {year}
  {2006})}\BibitemShut {NoStop}%
\bibitem [{\citenamefont {Borrero-Echeverry}\ \emph
  {et~al.}({2010})\citenamefont {Borrero-Echeverry}, \citenamefont {Schatz},\
  and\ \citenamefont {Tagg}}]{bor10}%
  \BibitemOpen
  \bibfield  {author} {\bibinfo {author} {\bibfnamefont {D.}~\bibnamefont
  {Borrero-Echeverry}}, \bibinfo {author} {\bibfnamefont {M.~F.}\ \bibnamefont
  {Schatz}}, \ and\ \bibinfo {author} {\bibfnamefont {R.}~\bibnamefont
  {Tagg}},\ }\bibfield  {title} {\enquote {\bibinfo {title} {{Transient
  turbulence in {{Taylor-Couette}} flow}},}\ }\href@noop {} {\bibfield
  {journal} {\bibinfo  {journal} {{Phys. Rev. E}}\ }\textbf {\bibinfo {volume}
  {{81}}},\ \bibinfo {pages} {025301} (\bibinfo {year} {{2010}})}\BibitemShut
  {NoStop}%
\bibitem [{\citenamefont {Ravelet}\ \emph {et~al.}(2010)\citenamefont
  {Ravelet}, \citenamefont {Delfos},\ and\ \citenamefont {Westerweel}}]{rav10}%
  \BibitemOpen
  \bibfield  {author} {\bibinfo {author} {\bibfnamefont {F.}~\bibnamefont
  {Ravelet}}, \bibinfo {author} {\bibfnamefont {R.}~\bibnamefont {Delfos}}, \
  and\ \bibinfo {author} {\bibfnamefont {J.}~\bibnamefont {Westerweel}},\
  }\bibfield  {title} {\enquote {\bibinfo {title} {Influence of global rotation
  and {{Reynolds}} number on the large-scale features of a turbulent
  {{Taylor--Couette}} flow},}\ }\href@noop {} {\bibfield  {journal} {\bibinfo
  {journal} {Phys. Fluids}\ }\textbf {\bibinfo {volume} {22}},\ \bibinfo
  {pages} {055103} (\bibinfo {year} {2010})}\BibitemShut {NoStop}%
\bibitem [{\citenamefont {Paoletti}\ and\ \citenamefont
  {Lathrop}(2011)}]{pao11}%
  \BibitemOpen
  \bibfield  {author} {\bibinfo {author} {\bibfnamefont {M.~S.}\ \bibnamefont
  {Paoletti}}\ and\ \bibinfo {author} {\bibfnamefont {D.~P.}\ \bibnamefont
  {Lathrop}},\ }\bibfield  {title} {\enquote {\bibinfo {title} {Angular
  momentum transport in turbulent flow between independently rotating
  cylinders},}\ }\href@noop {} {\bibfield  {journal} {\bibinfo  {journal}
  {Phys. Rev. Lett.}\ }\textbf {\bibinfo {volume} {106}},\ \bibinfo {pages}
  {024501} (\bibinfo {year} {2011})}\BibitemShut {NoStop}%
\bibitem [{\citenamefont {van Hout}\ and\ \citenamefont {Katz}(2011)}]{hou11}%
  \BibitemOpen
  \bibfield  {author} {\bibinfo {author} {\bibfnamefont {R.}~\bibnamefont {van
  Hout}}\ and\ \bibinfo {author} {\bibfnamefont {J.}~\bibnamefont {Katz}},\
  }\bibfield  {title} {\enquote {\bibinfo {title} {{Measurements of mean flow
  and turbulence characteristics in high-{{Reynolds}} number counter-rotating
  {{Taylor-Couette}} flow}},}\ }\href@noop {} {\bibfield  {journal} {\bibinfo
  {journal} {Phys. Fluids}\ }\textbf {\bibinfo {volume} {23}},\ \bibinfo
  {pages} {105102} (\bibinfo {year} {2011})}\BibitemShut {NoStop}%
\bibitem [{\citenamefont {van Gils}\ \emph {et~al.}(2012)\citenamefont {van
  Gils}, \citenamefont {Huisman}, \citenamefont {Grossmann}, \citenamefont
  {Sun},\ and\ \citenamefont {Lohse}}]{gil12}%
  \BibitemOpen
  \bibfield  {author} {\bibinfo {author} {\bibfnamefont {D.~P.~M.}\
  \bibnamefont {van Gils}}, \bibinfo {author} {\bibfnamefont {S.~G.}\
  \bibnamefont {Huisman}}, \bibinfo {author} {\bibfnamefont {S.}~\bibnamefont
  {Grossmann}}, \bibinfo {author} {\bibfnamefont {C.}~\bibnamefont {Sun}}, \
  and\ \bibinfo {author} {\bibfnamefont {D.}~\bibnamefont {Lohse}},\ }\bibfield
   {title} {\enquote {\bibinfo {title} {Optimal {{Taylor-Couette}}
  turbulence},}\ }\href@noop {} {\bibfield  {journal} {\bibinfo  {journal} {J.
  Fluid Mech.}\ }\textbf {\bibinfo {volume} {706}},\ \bibinfo {pages}
  {118--149} (\bibinfo {year} {2012})}\BibitemShut {NoStop}%
\bibitem [{\citenamefont {Huisman}\ \emph
  {et~al.}(2012{\natexlab{a}})\citenamefont {Huisman}, \citenamefont {van
  Gils}, \citenamefont {Grossmann}, \citenamefont {Sun},\ and\ \citenamefont
  {Lohse}}]{hui12}%
  \BibitemOpen
  \bibfield  {author} {\bibinfo {author} {\bibfnamefont {S.~G.}\ \bibnamefont
  {Huisman}}, \bibinfo {author} {\bibfnamefont {D.~P.~M.}\ \bibnamefont {van
  Gils}}, \bibinfo {author} {\bibfnamefont {S.}~\bibnamefont {Grossmann}},
  \bibinfo {author} {\bibfnamefont {C.}~\bibnamefont {Sun}}, \ and\ \bibinfo
  {author} {\bibfnamefont {D.}~\bibnamefont {Lohse}},\ }\bibfield  {title}
  {\enquote {\bibinfo {title} {Ultimate turbulent {{Taylor-Couette}} flow},}\
  }\href@noop {} {\bibfield  {journal} {\bibinfo  {journal} {Phys. Rev. Lett.}\
  }\textbf {\bibinfo {volume} {108}},\ \bibinfo {pages} {024501} (\bibinfo
  {year} {2012}{\natexlab{a}})}\BibitemShut {NoStop}%
\bibitem [{\citenamefont {Huisman}\ \emph {et~al.}(2014)\citenamefont
  {Huisman}, \citenamefont {van~der Veen}, \citenamefont {Sun},\ and\
  \citenamefont {Lohse}}]{hui14}%
  \BibitemOpen
  \bibfield  {author} {\bibinfo {author} {\bibfnamefont {S.~G.}\ \bibnamefont
  {Huisman}}, \bibinfo {author} {\bibfnamefont {R.~C.~A.}\ \bibnamefont
  {van~der Veen}}, \bibinfo {author} {\bibfnamefont {C.}~\bibnamefont {Sun}}, \
  and\ \bibinfo {author} {\bibfnamefont {D.}~\bibnamefont {Lohse}},\ }\bibfield
   {title} {\enquote {\bibinfo {title} {Multiple states in highly turbulent
  {{Taylor-Couette}} flow},}\ }\href@noop {} {\bibfield  {journal} {\bibinfo
  {journal} {Nat. Commun.}\ }\textbf {\bibinfo {volume} {5}},\ \bibinfo {pages}
  {3820} (\bibinfo {year} {2014})}\BibitemShut {NoStop}%
\bibitem [{\citenamefont {Schartman}\ \emph {et~al.}(2012)\citenamefont
  {Schartman}, \citenamefont {Ji}, \citenamefont {Burin},\ and\ \citenamefont
  {Goodman}}]{schartman12}%
  \BibitemOpen
  \bibfield  {author} {\bibinfo {author} {\bibfnamefont {E.}~\bibnamefont
  {Schartman}}, \bibinfo {author} {\bibfnamefont {Hantao}\ \bibnamefont {Ji}},
  \bibinfo {author} {\bibfnamefont {M.~J.}\ \bibnamefont {Burin}}, \ and\
  \bibinfo {author} {\bibfnamefont {J.}~\bibnamefont {Goodman}},\ }\bibfield
  {title} {\enquote {\bibinfo {title} {Stability of quasi-keplerian shear flow
  in a laboratory experiment},}\ }\href@noop {} {\bibfield  {journal} {\bibinfo
   {journal} {Astron. \& Astrophys.}\ }\textbf {\bibinfo {volume} {543}},\
  \bibinfo {pages} {A94} (\bibinfo {year} {2012})}\BibitemShut {NoStop}%
\bibitem [{\citenamefont {Merbold}\ \emph {et~al.}(2013)\citenamefont
  {Merbold}, \citenamefont {Brauckmann},\ and\ \citenamefont {Egbers}}]{mer13}%
  \BibitemOpen
  \bibfield  {author} {\bibinfo {author} {\bibfnamefont {S.}~\bibnamefont
  {Merbold}}, \bibinfo {author} {\bibfnamefont {H.~J.}\ \bibnamefont
  {Brauckmann}}, \ and\ \bibinfo {author} {\bibfnamefont {C.}~\bibnamefont
  {Egbers}},\ }\bibfield  {title} {\enquote {\bibinfo {title} {Torque
  measurements and numerical determination in differentially rotating wide gap
  {{{Taylor-Couette}}} flow},}\ }\href@noop {} {\bibfield  {journal} {\bibinfo
  {journal} {Phys. Rev. E}\ }\textbf {\bibinfo {volume} {87}},\ \bibinfo
  {pages} {023014} (\bibinfo {year} {2013})}\BibitemShut {NoStop}%
\bibitem [{\citenamefont {Bilson}\ and\ \citenamefont
  {Bremhorst}({2007})}]{bil07}%
  \BibitemOpen
  \bibfield  {author} {\bibinfo {author} {\bibfnamefont {M.}~\bibnamefont
  {Bilson}}\ and\ \bibinfo {author} {\bibfnamefont {K.}~\bibnamefont
  {Bremhorst}},\ }\bibfield  {title} {\enquote {\bibinfo {title} {{Direct
  numerical simulation of turbulent {{Taylor-Couette}} flow}},}\ }\href@noop {}
  {\bibfield  {journal} {\bibinfo  {journal} {{J. Fluid Mech.}}\ }\textbf
  {\bibinfo {volume} {{579}}},\ \bibinfo {pages} {227--270} (\bibinfo {year}
  {{2007}})}\BibitemShut {NoStop}%
\bibitem [{\citenamefont {He}\ \emph {et~al.}(2007)\citenamefont {He},
  \citenamefont {Tanahashi},\ and\ \citenamefont {Miyauchi}}]{heTC07}%
  \BibitemOpen
  \bibfield  {author} {\bibinfo {author} {\bibfnamefont {W.}~\bibnamefont
  {He}}, \bibinfo {author} {\bibfnamefont {M.}~\bibnamefont {Tanahashi}}, \
  and\ \bibinfo {author} {\bibfnamefont {T.}~\bibnamefont {Miyauchi}},\
  }\bibfield  {title} {\enquote {\bibinfo {title} {Direct numerical simulation
  of turbulent {{Taylor-Couette}} flow with high {{Reynolds}} number},}\
  }\href@noop {} {\bibfield  {journal} {\bibinfo  {journal} {Advances in
  Turbulence XI: Proceedings of the 11th EUROMECH European Turbulence
  Conference held June 25-28, 2007, in Porto, Portugal}\ } (\bibinfo {year}
  {2007})}\BibitemShut {NoStop}%
\bibitem [{\citenamefont {Dong}(2007)}]{don07}%
  \BibitemOpen
  \bibfield  {author} {\bibinfo {author} {\bibfnamefont {S.}~\bibnamefont
  {Dong}},\ }\bibfield  {title} {\enquote {\bibinfo {title} {Direct numerical
  simulation of turbulent {{Taylor-Couette}} flow},}\ }\href@noop {} {\bibfield
   {journal} {\bibinfo  {journal} {J. Fluid Mech.}\ }\textbf {\bibinfo {volume}
  {587}},\ \bibinfo {pages} {373--393} (\bibinfo {year} {2007})}\BibitemShut
  {NoStop}%
\bibitem [{\citenamefont {Pirro}\ and\ \citenamefont
  {Quadrio}({2008})}]{pir08}%
  \BibitemOpen
  \bibfield  {author} {\bibinfo {author} {\bibfnamefont {D.}~\bibnamefont
  {Pirro}}\ and\ \bibinfo {author} {\bibfnamefont {M.}~\bibnamefont
  {Quadrio}},\ }\bibfield  {title} {\enquote {\bibinfo {title} {{Direct
  numerical simulation of turbulent {{Taylor-Couette}} flow}},}\ }\href@noop {}
  {\bibfield  {journal} {\bibinfo  {journal} {{Eur. J. Mech. B-Fluids}}\
  }\textbf {\bibinfo {volume} {{27}}},\ \bibinfo {pages} {552--566} (\bibinfo
  {year} {{2008}})}\BibitemShut {NoStop}%
\bibitem [{\citenamefont {Brauckmann}\ and\ \citenamefont
  {Eckhardt}(2013)}]{bra13}%
  \BibitemOpen
  \bibfield  {author} {\bibinfo {author} {\bibfnamefont {H.~J.}\ \bibnamefont
  {Brauckmann}}\ and\ \bibinfo {author} {\bibfnamefont {B.}~\bibnamefont
  {Eckhardt}},\ }\bibfield  {title} {\enquote {\bibinfo {title} {Direct
  numerical simulations of local and global torque in {{Taylor-Couette}} flow
  up to {{Re}} = 30 000},}\ }\href@noop {} {\bibfield  {journal} {\bibinfo
  {journal} {J. Fluid Mech.}\ }\textbf {\bibinfo {volume} {718}},\ \bibinfo
  {pages} {398--427} (\bibinfo {year} {2013})}\BibitemShut {NoStop}%
\bibitem [{\citenamefont {Ostilla-M\'onico}\ \emph {et~al.}(2013)\citenamefont
  {Ostilla-M\'onico}, \citenamefont {Stevens}, \citenamefont {Grossmann},
  \citenamefont {Verzicco},\ and\ \citenamefont {Lohse}}]{ost13}%
  \BibitemOpen
  \bibfield  {author} {\bibinfo {author} {\bibfnamefont {R.}~\bibnamefont
  {Ostilla-M\'onico}}, \bibinfo {author} {\bibfnamefont {R.~J. A.~M.}\
  \bibnamefont {Stevens}}, \bibinfo {author} {\bibfnamefont {S.}~\bibnamefont
  {Grossmann}}, \bibinfo {author} {\bibfnamefont {R.}~\bibnamefont {Verzicco}},
  \ and\ \bibinfo {author} {\bibfnamefont {D.}~\bibnamefont {Lohse}},\
  }\bibfield  {title} {\enquote {\bibinfo {title} {Optimal {{Taylor-Couette}}
  flow: direct numerical simulations},}\ }\href@noop {} {\bibfield  {journal}
  {\bibinfo  {journal} {J. Fluid Mech.}\ }\textbf {\bibinfo {volume} {719}},\
  \bibinfo {pages} {14--46} (\bibinfo {year} {2013})}\BibitemShut {NoStop}%
\bibitem [{\citenamefont {Chouippe}\ \emph {et~al.}(2014)\citenamefont
  {Chouippe}, \citenamefont {Climent}, \citenamefont {Legendre},\ and\
  \citenamefont {Gabillet}}]{cho14}%
  \BibitemOpen
  \bibfield  {author} {\bibinfo {author} {\bibfnamefont {A.}~\bibnamefont
  {Chouippe}}, \bibinfo {author} {\bibfnamefont {E.}~\bibnamefont {Climent}},
  \bibinfo {author} {\bibfnamefont {D.}~\bibnamefont {Legendre}}, \ and\
  \bibinfo {author} {\bibfnamefont {C.}~\bibnamefont {Gabillet}},\ }\bibfield
  {title} {\enquote {\bibinfo {title} {Numerical simulation of bubble
  dispersion in turbulent {{Taylor-Couette}} flow},}\ }\href@noop {} {\bibfield
   {journal} {\bibinfo  {journal} {Phys. Fluids}\ }\textbf {\bibinfo {volume}
  {26}},\ \bibinfo {eid} {043304} (\bibinfo {year} {2014})}\BibitemShut
  {NoStop}%
\bibitem [{\citenamefont {Grossmann}\ and\ \citenamefont
  {Lohse}(2000)}]{gro00}%
  \BibitemOpen
  \bibfield  {author} {\bibinfo {author} {\bibfnamefont {S.}~\bibnamefont
  {Grossmann}}\ and\ \bibinfo {author} {\bibfnamefont {D.}~\bibnamefont
  {Lohse}},\ }\bibfield  {title} {\enquote {\bibinfo {title} {Scaling in
  thermal convection: A unifying theory},}\ }\href@noop {} {\bibfield
  {journal} {\bibinfo  {journal} {J. Fluid. Mech.}\ }\textbf {\bibinfo {volume}
  {407}},\ \bibinfo {pages} {27--56} (\bibinfo {year} {2000})}\BibitemShut
  {NoStop}%
\bibitem [{\citenamefont {Eckhardt}\ \emph {et~al.}(2007)\citenamefont
  {Eckhardt}, \citenamefont {Grossmann},\ and\ \citenamefont {Lohse}}]{eck07b}%
  \BibitemOpen
  \bibfield  {author} {\bibinfo {author} {\bibfnamefont {B.}~\bibnamefont
  {Eckhardt}}, \bibinfo {author} {\bibfnamefont {S.}~\bibnamefont {Grossmann}},
  \ and\ \bibinfo {author} {\bibfnamefont {D.}~\bibnamefont {Lohse}},\
  }\bibfield  {title} {\enquote {\bibinfo {title} {Torque scaling in turbulent
  {{Taylor-Couette}} flow between independently rotating cylinders},}\
  }\href@noop {} {\bibfield  {journal} {\bibinfo  {journal} {J. Fluid Mech.}\
  }\textbf {\bibinfo {volume} {581}},\ \bibinfo {pages} {221--250} (\bibinfo
  {year} {2007})}\BibitemShut {NoStop}%
\bibitem [{\citenamefont {Fenstermacher}\ \emph {et~al.}(1979)\citenamefont
  {Fenstermacher}, \citenamefont {Swinney},\ and\ \citenamefont
  {Gollub}}]{fen79}%
  \BibitemOpen
  \bibfield  {author} {\bibinfo {author} {\bibfnamefont {P.~R.}\ \bibnamefont
  {Fenstermacher}}, \bibinfo {author} {\bibfnamefont {Harry~L.}\ \bibnamefont
  {Swinney}}, \ and\ \bibinfo {author} {\bibfnamefont {J.~P.}\ \bibnamefont
  {Gollub}},\ }\bibfield  {title} {\enquote {\bibinfo {title} {Dynamical
  instabilities and the transition to chaotic {{{Taylor}}} vortex flow},}\
  }\href@noop {} {\bibfield  {journal} {\bibinfo  {journal} {J. Fluid Mech.}\
  }\textbf {\bibinfo {volume} {94}},\ \bibinfo {pages} {103--128} (\bibinfo
  {year} {1979})}\BibitemShut {NoStop}%
\bibitem [{\citenamefont {Andereck}\ \emph {et~al.}(1986)\citenamefont
  {Andereck}, \citenamefont {Liu},\ and\ \citenamefont {Swinney}}]{and86}%
  \BibitemOpen
  \bibfield  {author} {\bibinfo {author} {\bibfnamefont {C.~D.}\ \bibnamefont
  {Andereck}}, \bibinfo {author} {\bibfnamefont {S.~S.}\ \bibnamefont {Liu}}, \
  and\ \bibinfo {author} {\bibfnamefont {H.~L.}\ \bibnamefont {Swinney}},\
  }\bibfield  {title} {\enquote {\bibinfo {title} {Flow regimes in a circular
  {{Couette}} system with independently rotating cylinders},}\ }\href@noop {}
  {\bibfield  {journal} {\bibinfo  {journal} {J. Fluid Mech.}\ }\textbf
  {\bibinfo {volume} {164}},\ \bibinfo {pages} {155--183} (\bibinfo {year}
  {1986})}\BibitemShut {NoStop}%
\bibitem [{\citenamefont {Tokgoz}\ \emph {et~al.}(2011)\citenamefont {Tokgoz},
  \citenamefont {Elsinga}, \citenamefont {Delfos},\ and\ \citenamefont
  {Westerweel}}]{tok11}%
  \BibitemOpen
  \bibfield  {author} {\bibinfo {author} {\bibfnamefont {S.}~\bibnamefont
  {Tokgoz}}, \bibinfo {author} {\bibfnamefont {G.~E.}\ \bibnamefont {Elsinga}},
  \bibinfo {author} {\bibfnamefont {R.}~\bibnamefont {Delfos}}, \ and\ \bibinfo
  {author} {\bibfnamefont {J.}~\bibnamefont {Westerweel}},\ }\bibfield  {title}
  {\enquote {\bibinfo {title} {Experimental investigation of torque scaling and
  coherent structures in turbulent {{Taylor-Couette flow}}},}\ }\href@noop {}
  {\bibfield  {journal} {\bibinfo  {journal} {J. Phys.: Conf. Ser.}\ }\textbf
  {\bibinfo {volume} {318}},\ \bibinfo {pages} {082018} (\bibinfo {year}
  {2011})}\BibitemShut {NoStop}%
\bibitem [{\citenamefont {Mart\'inez-Arias}\ \emph {et~al.}(2014)\citenamefont
  {Mart\'inez-Arias}, \citenamefont {Peixinho}, \citenamefont {Crumeyrolle},\
  and\ \citenamefont {Mutabazi}}]{mar14}%
  \BibitemOpen
  \bibfield  {author} {\bibinfo {author} {\bibfnamefont {B.}~\bibnamefont
  {Mart\'inez-Arias}}, \bibinfo {author} {\bibfnamefont {J.}~\bibnamefont
  {Peixinho}}, \bibinfo {author} {\bibfnamefont {O.}~\bibnamefont
  {Crumeyrolle}}, \ and\ \bibinfo {author} {\bibfnamefont {I.}~\bibnamefont
  {Mutabazi}},\ }\bibfield  {title} {\enquote {\bibinfo {title} {Effect of the
  number of vortices on the torque scaling in {{Taylor-Couette}} flow},}\
  }\href@noop {} {\bibfield  {journal} {\bibinfo  {journal} {J. Fluid Mech.}\
  }\textbf {\bibinfo {volume} {748}},\ \bibinfo {pages} {756--767} (\bibinfo
  {year} {2014})}\BibitemShut {NoStop}%
\bibitem [{\citenamefont {Ostilla-M\'onico}\ \emph {et~al.}(2014)\citenamefont
  {Ostilla-M\'onico}, \citenamefont {van~der Poel}, \citenamefont {Verzicco},
  \citenamefont {Grossmann},\ and\ \citenamefont {Lohse}}]{ost14pd}%
  \BibitemOpen
  \bibfield  {author} {\bibinfo {author} {\bibfnamefont {R.}~\bibnamefont
  {Ostilla-M\'onico}}, \bibinfo {author} {\bibfnamefont {E.~P.}\ \bibnamefont
  {van~der Poel}}, \bibinfo {author} {\bibfnamefont {R.}~\bibnamefont
  {Verzicco}}, \bibinfo {author} {\bibfnamefont {S.}~\bibnamefont {Grossmann}},
  \ and\ \bibinfo {author} {\bibfnamefont {D.}~\bibnamefont {Lohse}},\
  }\bibfield  {title} {\enquote {\bibinfo {title} {Exploring the phase diagram
  of fully turbulent {{Taylor-Couette}} flow},}\ }\href@noop {} {\bibfield
  {journal} {\bibinfo  {journal} {J. Fluid Mech.}\ }\textbf {\bibinfo {volume}
  {761}},\ \bibinfo {pages} {1--26} (\bibinfo {year} {2014})}\BibitemShut
  {NoStop}%
\bibitem [{\citenamefont {Kraichnan}(1962)}]{kra62}%
  \BibitemOpen
  \bibfield  {author} {\bibinfo {author} {\bibfnamefont {R.~H.}\ \bibnamefont
  {Kraichnan}},\ }\bibfield  {title} {\enquote {\bibinfo {title} {Turbulent
  thermal convection at arbritrary {{Prandtl}} number},}\ }\href@noop {}
  {\bibfield  {journal} {\bibinfo  {journal} {Phys. Fluids}\ }\textbf {\bibinfo
  {volume} {5}},\ \bibinfo {pages} {1374--1389} (\bibinfo {year}
  {1962})}\BibitemShut {NoStop}%
\bibitem [{\citenamefont {Grossmann}\ and\ \citenamefont
  {Lohse}(2011)}]{gro11}%
  \BibitemOpen
  \bibfield  {author} {\bibinfo {author} {\bibfnamefont {S.}~\bibnamefont
  {Grossmann}}\ and\ \bibinfo {author} {\bibfnamefont {D.}~\bibnamefont
  {Lohse}},\ }\bibfield  {title} {\enquote {\bibinfo {title} {Multiple scaling
  in the ultimate regime of thermal convection},}\ }\href@noop {} {\bibfield
  {journal} {\bibinfo  {journal} {Phys. Fluids}\ }\textbf {\bibinfo {volume}
  {23}},\ \bibinfo {pages} {045108} (\bibinfo {year} {2011})}\BibitemShut
  {NoStop}%
\bibitem [{\citenamefont {He}\ \emph {et~al.}(2012)\citenamefont {He},
  \citenamefont {Funfschilling}, \citenamefont {Nobach}, \citenamefont
  {Bodenschatz},\ and\ \citenamefont {Ahlers}}]{he12}%
  \BibitemOpen
  \bibfield  {author} {\bibinfo {author} {\bibfnamefont {X.}~\bibnamefont
  {He}}, \bibinfo {author} {\bibfnamefont {D.}~\bibnamefont {Funfschilling}},
  \bibinfo {author} {\bibfnamefont {H.}~\bibnamefont {Nobach}}, \bibinfo
  {author} {\bibfnamefont {E.}~\bibnamefont {Bodenschatz}}, \ and\ \bibinfo
  {author} {\bibfnamefont {G.}~\bibnamefont {Ahlers}},\ }\bibfield  {title}
  {\enquote {\bibinfo {title} {Transition to the ultimate state of turbulent
  {{Rayleigh-B\'enard}} convection},}\ }\href@noop {} {\bibfield  {journal}
  {\bibinfo  {journal} {Phys. Rev. Lett.}\ }\textbf {\bibinfo {volume} {108}},\
  \bibinfo {pages} {024502} (\bibinfo {year} {2012})}\BibitemShut {NoStop}%
\bibitem [{\citenamefont {Kolmogorov}(1941{\natexlab{a}})}]{kol41a}%
  \BibitemOpen
  \bibfield  {author} {\bibinfo {author} {\bibfnamefont {A.~N.}\ \bibnamefont
  {Kolmogorov}},\ }\bibfield  {title} {\enquote {\bibinfo {title} {The local
  structure of turbulence in incompressible viscous fluid for very large
  {{Reynolds}} numbers},}\ }\href@noop {} {\bibfield  {journal} {\bibinfo
  {journal} {Dokl. Akad. Nauk SSSR}\ }\textbf {\bibinfo {volume} {30}},\
  \bibinfo {pages} {9--13} (\bibinfo {year} {1941}{\natexlab{a}})},\ \bibinfo
  {note} {reprinted in Proc. R. Soc. Lond. A {\bf 434}, 9-13
  (1991)}\BibitemShut {NoStop}%
\bibitem [{\citenamefont {Kolmogorov}(1941{\natexlab{b}})}]{kol41b}%
  \BibitemOpen
  \bibfield  {author} {\bibinfo {author} {\bibfnamefont {A.~N.}\ \bibnamefont
  {Kolmogorov}},\ }\bibfield  {title} {\enquote {\bibinfo {title} {On
  degeneration (decay) of isotropic turbulence in incompressible viscous
  liquid},}\ }\href@noop {} {\bibfield  {journal} {\bibinfo  {journal} {Dokl.
  Akad. Nauk SSSR}\ }\textbf {\bibinfo {volume} {31}},\ \bibinfo {pages}
  {538--540} (\bibinfo {year} {1941}{\natexlab{b}})}\BibitemShut {NoStop}%
\bibitem [{\citenamefont {Xi}\ and\ \citenamefont {Xia}(2008)}]{xi08}%
  \BibitemOpen
  \bibfield  {author} {\bibinfo {author} {\bibfnamefont {H.-D.}\ \bibnamefont
  {Xi}}\ and\ \bibinfo {author} {\bibfnamefont {K.-Q.}\ \bibnamefont {Xia}},\
  }\bibfield  {title} {\enquote {\bibinfo {title} {Flow mode transitions in
  turbulent thermal convection},}\ }\href@noop {} {\bibfield  {journal}
  {\bibinfo  {journal} {Phys. Fluids}\ }\textbf {\bibinfo {volume} {20}},\
  \bibinfo {pages} {055104} (\bibinfo {year} {2008})}\BibitemShut {NoStop}%
\bibitem [{\citenamefont {van~der Poel}\ \emph {et~al.}(2011)\citenamefont
  {van~der Poel}, \citenamefont {Stevens},\ and\ \citenamefont
  {Lohse}}]{poe11}%
  \BibitemOpen
  \bibfield  {author} {\bibinfo {author} {\bibfnamefont {E.~P.}\ \bibnamefont
  {van~der Poel}}, \bibinfo {author} {\bibfnamefont {R.~J. A.~M.}\ \bibnamefont
  {Stevens}}, \ and\ \bibinfo {author} {\bibfnamefont {D.}~\bibnamefont
  {Lohse}},\ }\bibfield  {title} {\enquote {\bibinfo {title} {Connecting flow
  structures and heat flux in turbulent {{Rayleigh-B\'enard}} convection},}\
  }\href@noop {} {\bibfield  {journal} {\bibinfo  {journal} {Phys. Rev. E}\
  }\textbf {\bibinfo {volume} {84}},\ \bibinfo {pages} {045303(R)} (\bibinfo
  {year} {2011})}\BibitemShut {NoStop}%
\bibitem [{\citenamefont {Weiss}\ and\ \citenamefont {Ahlers}(2013)}]{wei13}%
  \BibitemOpen
  \bibfield  {author} {\bibinfo {author} {\bibfnamefont {S.}~\bibnamefont
  {Weiss}}\ and\ \bibinfo {author} {\bibfnamefont {G.}~\bibnamefont {Ahlers}},\
  }\bibfield  {title} {\enquote {\bibinfo {title} {Effect of tilting on
  turbulent convection: cylindrical samples with aspect ratio
  {{$\Gamma$=0.50}}},}\ }\href@noop {} {\bibfield  {journal} {\bibinfo
  {journal} {J. Fluid. Mech.}\ }\textbf {\bibinfo {volume} {715}},\ \bibinfo
  {pages} {314--334} (\bibinfo {year} {2013})}\BibitemShut {NoStop}%
\bibitem [{\citenamefont {Ahlers}\ \emph {et~al.}(2011)\citenamefont {Ahlers},
  \citenamefont {Funfschilling},\ and\ \citenamefont
  {Bodenschatz}}]{RBchimney}%
  \BibitemOpen
  \bibfield  {author} {\bibinfo {author} {\bibfnamefont {G.}~\bibnamefont
  {Ahlers}}, \bibinfo {author} {\bibfnamefont {D.}~\bibnamefont
  {Funfschilling}}, \ and\ \bibinfo {author} {\bibfnamefont {E.}~\bibnamefont
  {Bodenschatz}},\ }\bibfield  {title} {\enquote {\bibinfo {title} {Heat
  transport in turbulent {{{Rayleigh-B\'enard}}} convection for {{{Pr}}}
  $\simeq$ 0.8 and {{{Ra}}} $\lesssim 10^{15}$},}\ }\href@noop {} {\bibfield
  {journal} {\bibinfo  {journal} {J. Phys.: Conf. Series}\ }\textbf {\bibinfo
  {volume} {318}},\ \bibinfo {pages} {082001} (\bibinfo {year}
  {2011})}\BibitemShut {NoStop}%
\bibitem [{\citenamefont {Wei}\ \emph {et~al.}(2015)\citenamefont {Wei},
  \citenamefont {Weiss},\ and\ \citenamefont {Ahlers}}]{wei15}%
  \BibitemOpen
  \bibfield  {author} {\bibinfo {author} {\bibfnamefont {P.}~\bibnamefont
  {Wei}}, \bibinfo {author} {\bibfnamefont {S.}~\bibnamefont {Weiss}}, \ and\
  \bibinfo {author} {\bibfnamefont {G.}~\bibnamefont {Ahlers}},\ }\bibfield
  {title} {\enquote {\bibinfo {title} {Multiple transitions in rotating
  turbulent rayleigh-b\'enard convection},}\ }\href@noop {} {\bibfield
  {journal} {\bibinfo  {journal} {Phys. Rev. Lett.}\ }\textbf {\bibinfo
  {volume} {114}},\ \bibinfo {pages} {114506} (\bibinfo {year}
  {2015})}\BibitemShut {NoStop}%
\bibitem [{\citenamefont {Ravelet}\ \emph {et~al.}(2004)\citenamefont
  {Ravelet}, \citenamefont {Mari\'e}, \citenamefont {Chiffaudel},\ and\
  \citenamefont {Daviaud}}]{Ravelet-PRL2004}%
  \BibitemOpen
  \bibfield  {author} {\bibinfo {author} {\bibfnamefont {F.}~\bibnamefont
  {Ravelet}}, \bibinfo {author} {\bibfnamefont {L.}~\bibnamefont {Mari\'e}},
  \bibinfo {author} {\bibfnamefont {A.}~\bibnamefont {Chiffaudel}}, \ and\
  \bibinfo {author} {\bibfnamefont {F.}~\bibnamefont {Daviaud}},\ }\bibfield
  {title} {\enquote {\bibinfo {title} {Multistability and memory effect in a
  highly turbulent flow: Experimental evidence for a global bifurcation},}\
  }\href@noop {} {\bibfield  {journal} {\bibinfo  {journal} {Phys. Rev. Lett.}\
  }\textbf {\bibinfo {volume} {93}},\ \bibinfo {pages} {164501} (\bibinfo
  {year} {2004})}\BibitemShut {NoStop}%
\bibitem [{\citenamefont {Ravelet}\ \emph {et~al.}(2008)\citenamefont
  {Ravelet}, \citenamefont {Chiffaudel},\ and\ \citenamefont
  {Daviaud}}]{Ravelet-JFM2008}%
  \BibitemOpen
  \bibfield  {author} {\bibinfo {author} {\bibfnamefont {F.}~\bibnamefont
  {Ravelet}}, \bibinfo {author} {\bibfnamefont {A.}~\bibnamefont {Chiffaudel}},
  \ and\ \bibinfo {author} {\bibfnamefont {F.}~\bibnamefont {Daviaud}},\
  }\bibfield  {title} {\enquote {\bibinfo {title} {Supercritical transition to
  turbulence in an inertially driven von {{K\'am\'an}} closed flow},}\
  }\href@noop {} {\bibfield  {journal} {\bibinfo  {journal} {J. Fluid Mech.}\
  }\textbf {\bibinfo {volume} {601}},\ \bibinfo {pages} {339--364} (\bibinfo
  {year} {2008})}\BibitemShut {NoStop}%
\bibitem [{\citenamefont {Cortet}\ \emph {et~al.}(2010)\citenamefont {Cortet},
  \citenamefont {Chiffaudel}, \citenamefont {Daviaud},\ and\ \citenamefont
  {Dubrulle}}]{Cortet-PRL2010}%
  \BibitemOpen
  \bibfield  {author} {\bibinfo {author} {\bibfnamefont {P.-P.}\ \bibnamefont
  {Cortet}}, \bibinfo {author} {\bibfnamefont {A.}~\bibnamefont {Chiffaudel}},
  \bibinfo {author} {\bibfnamefont {F.}~\bibnamefont {Daviaud}}, \ and\
  \bibinfo {author} {\bibfnamefont {B.}~\bibnamefont {Dubrulle}},\ }\bibfield
  {title} {\enquote {\bibinfo {title} {Experimental evidence of a phase
  transition in a closed turbulent flow},}\ }\href@noop {} {\bibfield
  {journal} {\bibinfo  {journal} {Phys. Rev. Lett.}\ }\textbf {\bibinfo
  {volume} {105}},\ \bibinfo {pages} {214501} (\bibinfo {year}
  {2010})}\BibitemShut {NoStop}%
\bibitem [{\citenamefont {Zimmerman}\ \emph {et~al.}(2011)\citenamefont
  {Zimmerman}, \citenamefont {Triana},\ and\ \citenamefont
  {Lathrop}}]{zimmerman2011}%
  \BibitemOpen
  \bibfield  {author} {\bibinfo {author} {\bibfnamefont {D.~S.}\ \bibnamefont
  {Zimmerman}}, \bibinfo {author} {\bibfnamefont {S.~A.}\ \bibnamefont
  {Triana}}, \ and\ \bibinfo {author} {\bibfnamefont {D.~P.}\ \bibnamefont
  {Lathrop}},\ }\bibfield  {title} {\enquote {\bibinfo {title} {Bi-stability in
  turbulent, rotating spherical {{{Couette}}} flow},}\ }\href@noop {}
  {\bibfield  {journal} {\bibinfo  {journal} {Phys. Fluids}\ }\textbf {\bibinfo
  {volume} {23}},\ \bibinfo {pages} {065104} (\bibinfo {year}
  {2011})}\BibitemShut {NoStop}%
\bibitem [{\citenamefont {Gul}\ \emph {et~al.}(2015)\citenamefont {Gul},
  \citenamefont {Elsinga},\ and\ \citenamefont {Westerweel}}]{gul15}%
  \BibitemOpen
  \bibfield  {author} {\bibinfo {author} {\bibfnamefont {M.}~\bibnamefont
  {Gul}}, \bibinfo {author} {\bibfnamefont {G.~E.}\ \bibnamefont {Elsinga}}, \
  and\ \bibinfo {author} {\bibfnamefont {J.}~\bibnamefont {Westerweel}},\
  }\bibfield  {title} {\enquote {\bibinfo {title} {Experimental investigation
  of geometry on torque hysteresis behaviour of taylor-couette flow},}\
  }\href@noop {} {\bibfield  {journal} {\bibinfo  {journal} {15th European
  Turbulence Conference held August 25-28, 2015, in Delft, The Netherlands}\ }
  (\bibinfo {year} {2015})}\BibitemShut {NoStop}%
\bibitem [{\citenamefont {van Gils}\ \emph
  {et~al.}(2011{\natexlab{a}})\citenamefont {van Gils}, \citenamefont
  {Bruggert}, \citenamefont {Lathrop}, \citenamefont {Sun},\ and\ \citenamefont
  {Lohse}}]{gil11a}%
  \BibitemOpen
  \bibfield  {author} {\bibinfo {author} {\bibfnamefont {D.~P.~M.}\
  \bibnamefont {van Gils}}, \bibinfo {author} {\bibfnamefont {G.~W.}\
  \bibnamefont {Bruggert}}, \bibinfo {author} {\bibfnamefont {D.~P.}\
  \bibnamefont {Lathrop}}, \bibinfo {author} {\bibfnamefont {C.}~\bibnamefont
  {Sun}}, \ and\ \bibinfo {author} {\bibfnamefont {D.}~\bibnamefont {Lohse}},\
  }\bibfield  {title} {\enquote {\bibinfo {title} {The {{Twente}} turbulent
  {{Taylor-Couette}} {{($T^3C$)}} facility: strongly turbulent (multi-phase)
  flow between independently rotating cylinders},}\ }\href@noop {} {\bibfield
  {journal} {\bibinfo  {journal} {Rev. Sci. Instr.}\ }\textbf {\bibinfo
  {volume} {82}},\ \bibinfo {pages} {025105} (\bibinfo {year}
  {2011}{\natexlab{a}})}\BibitemShut {NoStop}%
\bibitem [{\citenamefont {Huisman}\ \emph {et~al.}(2015)\citenamefont
  {Huisman}, \citenamefont {van~der Veen}, \citenamefont {Bruggert},
  \citenamefont {Lohse},\ and\ \citenamefont {Sun}}]{hui15}%
  \BibitemOpen
  \bibfield  {author} {\bibinfo {author} {\bibfnamefont {S.~G.}\ \bibnamefont
  {Huisman}}, \bibinfo {author} {\bibfnamefont {R.~C.~A.}\ \bibnamefont
  {van~der Veen}}, \bibinfo {author} {\bibfnamefont {G.~W.}\ \bibnamefont
  {Bruggert}}, \bibinfo {author} {\bibfnamefont {D.}~\bibnamefont {Lohse}}, \
  and\ \bibinfo {author} {\bibfnamefont {C.}~\bibnamefont {Sun}},\ }\bibfield
  {title} {\enquote {\bibinfo {title} {The boiling {{Twente}}
  {{Taylor-Couette}} {{(BTTC)}} facility: Temperature controlled turbulent flow
  between independently rotating, coaxial cylinders},}\ }\href@noop {}
  {\bibfield  {journal} {\bibinfo  {journal} {Rev. Sci. Instr.}\ }\textbf
  {\bibinfo {volume} {86}},\ \bibinfo {pages} {065108} (\bibinfo {year}
  {2015})}\BibitemShut {NoStop}%
\bibitem [{\citenamefont {Huisman}\ \emph
  {et~al.}(2012{\natexlab{b}})\citenamefont {Huisman}, \citenamefont {van
  Gils},\ and\ \citenamefont {Sun}}]{hui12b}%
  \BibitemOpen
  \bibfield  {author} {\bibinfo {author} {\bibfnamefont {S.~G.}\ \bibnamefont
  {Huisman}}, \bibinfo {author} {\bibfnamefont {D.~P.~M.}\ \bibnamefont {van
  Gils}}, \ and\ \bibinfo {author} {\bibfnamefont {C.}~\bibnamefont {Sun}},\
  }\bibfield  {title} {\enquote {\bibinfo {title} {Applying laser {{Doppler}}
  anemometry inside a {{Taylor-Couette}} geometry using a ray-tracer to correct
  for curvature effects},}\ }\href@noop {} {\bibfield  {journal} {\bibinfo
  {journal} {Eur. J. Mech. B/Fluids}\ }\textbf {\bibinfo {volume} {36}},\
  \bibinfo {pages} {115--119} (\bibinfo {year}
  {2012}{\natexlab{b}})}\BibitemShut {NoStop}%
\bibitem [{\citenamefont {van Gils}\ \emph
  {et~al.}(2011{\natexlab{b}})\citenamefont {van Gils}, \citenamefont
  {Huisman}, \citenamefont {Bruggert}, \citenamefont {Sun},\ and\ \citenamefont
  {Lohse}}]{gil11}%
  \BibitemOpen
  \bibfield  {author} {\bibinfo {author} {\bibfnamefont {D.~P.~M.}\
  \bibnamefont {van Gils}}, \bibinfo {author} {\bibfnamefont {S.~G.}\
  \bibnamefont {Huisman}}, \bibinfo {author} {\bibfnamefont {G.~W.}\
  \bibnamefont {Bruggert}}, \bibinfo {author} {\bibfnamefont {C.}~\bibnamefont
  {Sun}}, \ and\ \bibinfo {author} {\bibfnamefont {D.}~\bibnamefont {Lohse}},\
  }\bibfield  {title} {\enquote {\bibinfo {title} {Torque scaling in turbulent
  {{Taylor-Couette}} flow with co- and counter-rotating cylinders},}\
  }\href@noop {} {\bibfield  {journal} {\bibinfo  {journal} {Phys. Rev. Lett.}\
  }\textbf {\bibinfo {volume} {106}},\ \bibinfo {pages} {024502} (\bibinfo
  {year} {2011}{\natexlab{b}})}\BibitemShut {NoStop}%
\bibitem [{\citenamefont {L\'opez-Caballero}\ and\ \citenamefont
  {Burguete}(2013)}]{caballero2013}%
  \BibitemOpen
  \bibfield  {author} {\bibinfo {author} {\bibfnamefont {M.}~\bibnamefont
  {L\'opez-Caballero}}\ and\ \bibinfo {author} {\bibfnamefont {J.}~\bibnamefont
  {Burguete}},\ }\bibfield  {title} {\enquote {\bibinfo {title} {Inverse
  cascades sustained by the transfer rate of angular momentum in a 3d turbulent
  flow},}\ }\href@noop {} {\bibfield  {journal} {\bibinfo  {journal} {Phys.
  Rev. Lett.}\ }\textbf {\bibinfo {volume} {110}},\ \bibinfo {pages} {124501}
  (\bibinfo {year} {2013})}\BibitemShut {NoStop}%
\end{thebibliography}%

\end{document}